\documentclass[12pt]{article}

\usepackage[dvips]{epsfig,graphicx, color}

\usepackage{times}
\usepackage{amsmath, amssymb}

\newenvironment{sciabstract}{%
\begin{quote} \bf}
{\end{quote}}

\newcounter{lastnote}

\title{Differential gene expression in \textit{Bacillus subtilis}}

\date{}

\begin{document}


\baselineskip12pt

\maketitle
%{\LARGE SpoIIAB allostery enables differential gene expression in \textit{Bacillus subtilis}} \\[1.5cm]
%{\LARGE The allosteric behaviour of the protein kinase SpoIIAB enables differential gene expression in \textit{Bacillus subtilis}} \\[1.5cm]
{\large Dagmar Iber$^1$, Joanna Clarkson$^2$, Michael D. Yudkin$^2$, Iain D. Campbell$^3$\\[1cm]
{\normalsize $^{1}$Mathematical Institute, Centre for Mathematical
Biology,University of Oxford, 24-29 St Giles}\\
{\normalsize Oxford OX1 3LB,UK;E-mail: iber@maths.ox.ac.uk}\\[1ex]
{\normalsize $^2$Microbiology Unit,$^{2,3}$Department of
Biochemistry, University of Oxford, South Parks Road, OX1 3QU,
Oxford, UK}}

\newpage

\begin{sciabstract}
Sporulation in \textit{Bacillus subtilis} serves as a paradigm for
the development of two different cell types (mother cell and
prespore) from a single cell. The mechanism by which the two
different developmental programs are initiated has been much studied
but is not well understood. With the help of existing and new
experimental results, a mathematical model has been developed that
reproduces all published \textit{in vitro} experiments and makes new
predictions about the properties of the system \textit{in vivo}.
\end{sciabstract}
\newpage

\section*{Introduction}
Multicellular organisms achieve cell differentiation by selective
expression of genes but the mechanisms that trigger such
differential expression are difficult to define. Single-celled
prokaryotes offer several experimental advantages for studies of
cell differentiation; a good example is spore formation in Bacillus
subtilis which is induced by nutritional stress. During sporulation
the single cell organism differentiates into two cell types, a
smaller forespore (prespore) and a larger mother cell, separated by
a septum. The two different developmental programs are initiated by
the transcription factor $\sigma^F$, which associates with the core
RNA polymerase in the prespore but not in the mother cell. In the
mother cell $\sigma^F$ is trapped in a complex with SpoIIAB
(hereafter referred to as AB). While the pathways that lead to
$\sigma^F$ dissociation are well-defined, it is not yet clear why
$\sigma^F$ association with the core RNA polymerase is restricted to
the prespore \cite{Hilbert04}.

Crystallography has demonstrated that one molecule of $\sigma^F$
binds asymmetrically across the two subunits of the AB dimer
\cite{Campbell02}. The binding of $\sigma^F$ prevents access of
SpoIIAA (hereafter referred to as AA) to one monomer of AB, but
allows docking of AA on the other monomer \cite{Ho03}. Steric and
electrostatic clashes are caused by the docking of AA, and
$\sigma^F$ is subsequently displaced from
 AB \cite{Masuda04,Ho03}. AA binds across the
nucleotide-binding pocket of AB, and is phosphorylated when this
binding pocket contains ATP, yielding inactive phosphorylated AA
(AA-P) and AB$\cdot$ADP. AB$\cdot$ADP can either be sequestered in
an AA$\cdot$AB$\cdot$ADP complex or exchange ADP for ATP. Free AB
(in both its ATP and ADP forms) can rebind $\sigma^F$, and sustained
$\sigma^F$ release is thus expected to require reactivation of AA-P
by SpoIIE-mediated dephosphorylation \cite{Magnin97}.

The restriction of $\sigma^F$ release to the prespore has been
ascribed to two aspects of asymmetric septation: i) the volume
difference between prespore and mother cell, together with the
accumulation of SpoIIE (hereafter referred to as IIE) on the septum
\cite{Duncan95}; ii) loss of AB in the prespore, caused by a
transient genetic imbalance, as described below
\cite{Frandsen99,Dworkin01}. Accumulation of IIE on (both sides of)
the septum results in an increase in IIE activity in the smaller
prespore and reduced activity in the larger mother cell. It has been
suggested that the increase in IIE activity may lead to sufficient
dephosphorylation of AA-P to trigger $\sigma^F$ release specifically
in the forespore \cite{Duncan95}, and recent biochemical experiments
indeed reveal that $\sigma^F \cdot$AB complexes are highly sensitive
to small changes in IIE activity \cite{Clarkson04_3}. The resulting
release of $\sigma^F$ in the prespore might be reinforced by rapid
degradation of unbound AB \cite{Pan01,Dworkin01}, which would
prevent reassociation of $\sigma^F$ and AB. Degradation of AB is not
balanced by protein expression in the prespore because of a
transient genetic imbalance; this arises from the fact that about
70$\%$ of the chromosome (including the \textit{spoIIA} operon in
which AB is coded) is initially excluded from the forespore and is
imported into it only 10-20 minutes after septum closure
\cite{Wu98}. A further proposed result of the genetic imbalance is
the degradation and loss of a IIE inhibitor from the prespore, which
would further accelerate $\sigma^F$ release \cite{Frandsen99}. The
nature of such an inhibitor is, however, still elusive, and while
the concentration of IIE on the septum and the degradation of AB in
the forespore have both been shown to be important for successful
sporulation, interference with either one of them does not abolish
sporulation \cite{Dworkin01,Arigoni99}. No model yet explains how
these small and transient differences act together to maintain the
release of $\sigma^F$ for at least 1h. Moreover, it is unclear
whether these processes are sufficient for the initiation of
sporulation or whether other, still  unidentified, factors  play a
role \cite{Hilbert04}.

By employing a combination of experimental and mathematical modeling
tools we derive the regulatory circuit that controls $\sigma^F$
release. A large amount of information is available about the
properties of this system, and most of the kinetic and thermodynamic
parameters are known. Our model quantitatively reproduces the
system's known behaviour \textit{in vitro}, including the
sensitivity of AB$\cdot\sigma^F$ complexes to small changes in IIE
activity and in AA concentration. The model predicts that SpoIIE
relocation to the septum is sufficient for prespore-specific
formation of  $\sigma^F \cdot$RNA polymerase holoenzyme and that the
required sensitivity stems from SpoIIAB allostery, which we confirm
experimentally. Stochastic fluctuations in gene expression are
buffered by coupled translation of SpoIIAA and SpoIIAB combined with
degradation of unbound SpoIIAB, while ATP consumption is limited by
the formation of SpoIIAA-SpoIIAB$\cdot$ADP. The transient genetic
imbalance plays no important role.

\section*{Construction of the mathematical model}
\subsection*{The pathway that controls $\sigma^F$ release}
The key steps and players involved in the regulation of $\sigma^F$
can be summarized as follows. $\sigma^F$ binds asymmetrically across
the two subunits of the AB dimer \cite{Campbell02} and is displaced
by steric and electrostatic clashes that are caused by the docking
of SpoIIAA (AA) \cite{Masuda04,Ho03}. AA binds across the
nucleotide-binding pocket of AB, and is phosphorylated when this
binding pocket contains ATP, yielding inactive phosphorylated AA
(AA-P) and AB$\cdot$ADP. AB$\cdot$ADP can either be sequestered in
an AA$\cdot$AB$\cdot$ADP complex or exchange ADP for ATP. Free AB
(in both its ATP and ADP forms) can rebind $\sigma^F$, and sustained
$\sigma^F$ release is thus expected to require reactivation of AA-P
by SpoIIE-mediated dephosphorylation \cite{Magnin97}.

\subsection*{A model for the pathway that controls $\sigma^F$ release}
During early attempts to develop a model of the system described in
Figure 1A in the main paper it became clear that some extensions
would be necessary. One important extension, confirmed by new
experiments is that AB, which is a dimer, binds AA with positive
cooperativity (see below). This cooperative behaviour increases the
number of reactions that need to be considered in the model. The
shaded part of the scheme summarizes the interactions of one AB
confomer and therefore includes only the well-known reactions that
have been reviewed extensively (see Hilbert and Piggot (2004) and
references therein). In brief, binding of $\sigma^F$ to AB$\cdot$ATP
(61/62 - all following numbers in round brackets refer to reactions
in our model) leads to the formation of a $\sigma^F
\cdot$AB$\cdot$ATP complex \cite{Duncan93}. $\sigma^F$ in this
complex cannot bind to core RNA polymerase and is thus termed
inactive \cite{Duncan93,Min93}. Subsequent interaction with AA leads
to the formation of $\sigma^F \cdot$AB$\cdot$ATP$\cdot$AA (43/44),
from which $\sigma^F$ is rapidly released (79/80)
\cite{Clarkson04_2}. Consistent with structural information derived
from crystallography \cite{Campbell02,Masuda04}, binding of two AA
to the AB dimer while $\sigma^F$ is still bound is impossible
\cite{Ho03}. The bound AA can be phosphorylated either immediately
(185) or after a further AA has bound (189) \cite{Min93}. AA-P can
be reativated by the phosphatase action of IIE (193-195;
\cite{Duncan95,Arigoni96}). AB$\cdot$ADP can either bind AA (9-12,
35/36,158) or rebind $\sigma^F$ - either immediately (63/64) or
after ADP-ATP exchange (140-143,170-173) \cite{Magnin97}. ADP-ATP
exchange is also possible when AA is bound to AB$\cdot$ADP (162,164)
and dimers with no, or two different, nucleotides bound
(19-26,53-56,69/70,86/87,92/93,156/157,178/179,186,191) also need to
be considered.

Given the allosteric behaviour of AB the scheme had to be extended
to include also a second AB confomer (non-shaded part) as well as
interconversions between AB states (94-137). This leads to a
duplication of almost all the interactions mentioned so far. Only in
the case of AA and $\sigma^F$ binding to nucleotide-free AB could
the unfavoured conformation be neglected, since a comparison of the
relevant parameters shows that these interactions would be of
significance neither in the \textit{in vitro} experiments nor in the
sporulating cell. In the diagram, the conformation that binds AA
with low affinity and ADP with higher affinity is depicted as
circles (AB), and the other state as squares ($\hat{AB}$). As is
usual for descriptions of allosteric proteins (e.g. \cite{Fersht})
these two states may be referred to as the tense (T) and the relaxed
(R) state respectively.

\subsubsection*{SpoIIAB is an allosteric protein}
Our experiments with fluorescence quenching, a Scatchard plot, and
surface plasmon resonance (SPR) all provide evidence for the
allosteric behaviour of SpoIIAB. The non-cooperative binding
mechanism of published models failed to fit results from
fluorescence quenching data obtained with AA and AB-F97W, a
fluorescent derivative of AB \cite{Clarkson04_1}. When AB-F97W was
mixed with AA, less than an equimolar AA concentration was needed to
quench half the fluorescence (Fig. \ref{Fig_allostery} A). The
effect was even stronger with the lower affinity AA mutant, AA-S58A,
suggesting that binding of one AA to AB induces a conformational
change which results in quenching of the fluorescence on the second
non-bound monomer. A Scatchard analysis confirmed the allosteric
behaviour of AB (see Fig. 2A in the main paper). The experiments had
to be carried out with the AA-S58A mutant to avoid AA
phosphorylation by trace amounts of ATP. (The AA/AB ratio is
apparently overestimated by a factor of about 1.3, presumably
because of errors in the determination of protein concentrations,
and these probably explain the $\nu>2$ result). Moreover, SPR
binding kinetics obtained with AB bound to the sensor chip are
better fitted by an allosteric model (see below). Consistent with
the hypothesis that AB is an allosteric protein, crystallographic
structures of AB in complex with either $\sigma^F$ or AA
\cite{Campbell02,Masuda04} have shown it to be a flexible molecule.
The conformation of AB is sensitive to the type of ligand bound,
with changes occurring within the nucleotide-binding region and in
the angle between the subunits. However, our model assumes that both
the $\sigma^F$- and AA-bound forms of AB are predominantly in the
R-state, with the result that changes between the states would not
be visible in the crystal structures because of low occupancy of the
T state.

\subsection*{Parameter determination}
Lack of quantitative data and experimental detail have often
restricted the predictive power of mathematical models. This study
has greatly benefited from the large amount of published
experimental information and the availability of a well-established
\textit{in vitro} system. Values for essentially all the key
parameters could thus be determined experimentally. The almost 200
rate constants included in the scheme were reduced to 30 independent
rate constants by assuming that the kinetic constants of one AB
monomer are not affected by ADP, ATP or AA being bound to the other
monomer other than through a change in the relative concentration of
the R and T states.

The 27 independent kinetic constants for the \textit{in vitro} model
can be grouped into those characterising the AB-AA interaction
(reactions 1-58), the AB-$\sigma^F$ interaction in the absence
(59-76) or presence (77-93) of AA, the interconversion between the R
and the T state (94-137), the ADP-ATP exchange (138-182), and
finally the phosphorylation (183-192) and dephosphorylation
(193-195) of AA. The values of many of these parameters have already
been published. However, most of the on- and off-rates have been
determined by SPR; this technique requires the proteins analysed to
be attached to a sensor chip, which may alter their behavior.
Fluorescence quenching experiments do not have this shortcoming, and
a re-analysis of important kinetic parameters was carried out by
fluorescence spectroscopy with the help of fluorescent mutants
\cite{Clarkson04_1,Clarkson04_2}. Constants that had been determined
on the assumption that AB is not an allosteric protein were refined
to fit the model.

\subsubsection*{AB-AA binding constants} The first group
of constants are those characterising the AB-AA interaction (1-58).
The rate constants for the formation of the AB$\cdot$AA complex in
the presence of ADP can be determined by SPR following the protocol
established by Magnin \textit{et al} (1997). These measurements can
be carried out with either AB or AA bound to the sensor chip.
However, AA-interactions with both AB monomers will occur only when
AB is bound to the chip, since the spacing of chip-bound AA is
unlikely to be appropriate for AB to bind to two molecules of AA.
For the monomeric interaction when AA is chip-bound a total of 15
experiments were pooled from Shu \textit{et al} (2004) together with
other unpublished results from this laboratory. The on- and
off-rates were determined as $k_{on}^* = 1.6 \times 10^5$
M$^{-1}$s$^{-1}$ ($\pm 2.1 \times 10^5$ standard deviation) and
$k_{off}^* = 6.85 \times 10^{-3}$ s$^{-1}$ ($\pm 4 \times 10^{-3}$
standard deviation) respectively. Here the off-rate will correspond
to the the high affinity state, since upon binding of AA, AB  will
undergo a conformational change to the high affinity form.

\begin{figure} [!b]
\begin{center}
\includegraphics[width=15cm]{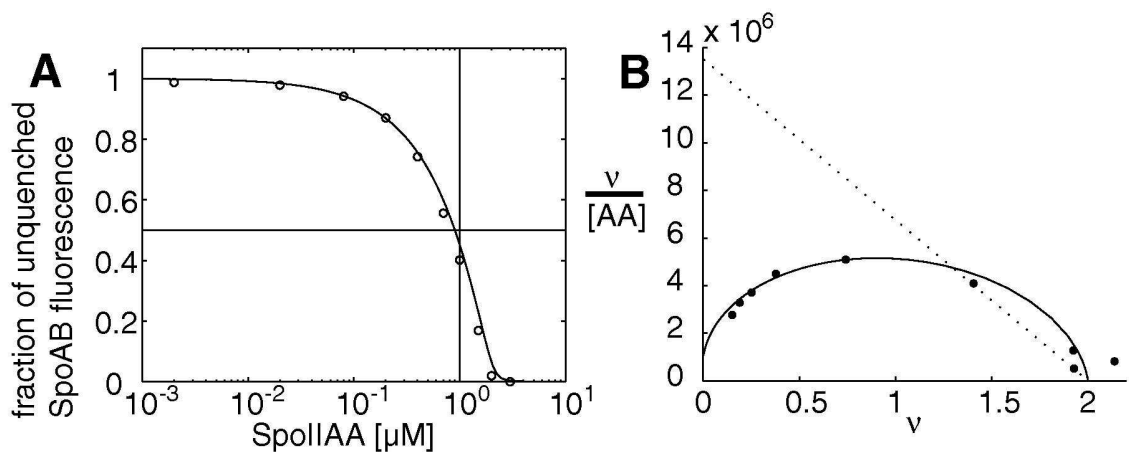} \caption
{\label{Fig_allostery} {\small \textbf{SpoIIAB is an allosteric
protein.} \textbf{(A)} Fraction of AB (1 $\mu$M total) unbound as a
function of the concentration of AA in the presence of 100 $\mu$M
ADP, as determined by fluorescence quenching experiments with
fluorescent mutant AB-F97W \cite{Clarkson04_1}. The points show
previously unpublished experimental results; the line represents the
binding curve expected from the kinetic constants employed in the
mathematical model (which includes allostery of AB). \textbf{(B)}
Scatchard plot obtained by determining the fraction of bound AB
($\nu$) when $4 \mu$M AA-S58A is mixed with different concentrations
of AB. The points show previously unpublished experimental results;
the lines represent the predicted Scatchard plot based on the
assumption that AB is an allosteric protein (continuous line) or
non-allosteric (dotted) with an affinity for AA-S58A$\cdot$AB 3.7
times less than for AB$\cdot$AA (Table \ref{Table_par}). Note that a
1.3-fold higher AB concentration than measured had to be used to
keep $\nu \leq 2$.}}
\end{center}
\end{figure}

To facilitate the fitting of the binding of two AA molecules to one
chip-bound AB molecule by means of the Biacore software we exploited
the fact that one conformation of the AB and AB$\cdot$AA$_2$ dimers
is particularly favored; we thus simplified the binding reaction to
a 3-step mechanism involving an initial binding of one AA to the AB
dimer (in the low affinity state), a subsequent conformational
switch and the binding of a second AA to AB$\cdot$AA in the high
affinity state. Fitting of SPR data with this 3-step mechanism was
substantially improved over that with a single step (non-allosteric)
model, with the average $\chi^2$ value of the fit being reduced by
approximately 10-fold. (In contrast fitting was not improved by the
3-step mechanism when AA was bound to the chip.) The estimates
obtained from 5 unpublished experiments for the on- and off-rates of
the two binding steps were $k_{on1} = 1.1 \times 10^6$
M$^{-1}$s$^{-1}$ ($\pm$ $10^6$ standard deviation), $k_{on2} = 4
\times 10^4$ M$^{-1}$s$^{-1}$ ($\pm 2.9 \times 10^4$ standard
deviation), $k_{off1} = 2.4$ s$^{-1}$ ($\pm 2.2$ standard
deviation), $k_{off2} = 8.4 \times 10^{-3}$ s$^{-1}$ ($\pm 1.7
\times 10^{-3}$ standard deviation). The conformational change
occurs at 7.4 s$^{-1}$ ($\pm 5.6$ standard deviation) and 2 s$^{-1}$
($\pm 1$ standard deviation) towards the high and low affinity
states respectively. The affinity of the second bound AA will
probably have been underestimated by SPR since attachment of AB to
the grid is likely to obscure at least one binding site and to
constrain conformational changes. Equally the AA-AB on-rate will
have been underestimated when AA is fixed to the grid since AA, the
smaller protein, will diffuse more rapidly when both are in
solution. We therefore used an intermediate value $k_{on1} = 8\times
10^5$ M$^{-1}$s$^{-1}$ as the general AA-AB on-rate in our
simulations, and $7.4\times 10^{-3}$ s$^{-1}$ as the off-rate in the
high affinity state, which is the average value of all measurements
made. $k_{off1}$ and the rate for the conformational change in the
disfavoured direction were used as determined by SPR above. In order
to fit AA-AB binding as measured in fluorescence quenching
experiments (Fig. \ref{Fig_allostery}) the conformational change
towards the favoured state had to be taken to be higher than
estimated by SPR. We used 20 s$^{-1}$; a higher value would further
improve the fit in Figure \ref{Fig_allostery}a. A likely explanation
for the lower value for the SPR measurement is that fixation to the
chip impairs AB conformational changes.

The binding constants in the presence of ATP are more difficult to
measure since AB$\cdot$ATP$\cdot$AA complexes are not stable because
of phosphorylation of AA under these conditions. SPR measurements of
the interaction of AB with the AA-S58A mutant of AA, which cannot be
phosphorylated by AB, revealed a higher affinity in the presence of
ATP than of ADP \cite{Magnin97}. However, the affinity of the
AA-S58A mutant in the presence of ADP is lower than for the
wildtype. We therefore could not directly use the values determined
for complexes of AA-S58A with AB$\cdot$ATP, and had to set the
second off-rate to a lower, arbitrarily chosen value ($k_{off2} =
0.001$ s$^{-1}$).

In the absence of nucleotides AA-AB binding is weak
\cite{Clarkson01} and we used an off-rate that corresponds to a
dissociation constant of 2.6 mM.

It was assumed that the presence of $\sigma^F$ did not  affect the
AB$\cdot$AA affinity \cite{Clarkson04_1}.

\subsubsection*{$\sigma^F$-AB binding constants}
The next group of constants for which SPR data have been published
\cite{Magnin97} are those that characterize the $\sigma^F$-AB
interaction (59-93). While these could be used for the
$\sigma^F$-AB$\cdot$ATP interaction, fluorescence spectroscopy data
suggest a lower $\sigma^F\cdot$AB$\cdot$ADP affinity than that
measured by SPR (Fig. \ref{Fig_par_det}A). Accordingly the
$\sigma^F\cdot$AB$\cdot$ADP off-rate was increased to $4$ s$^{-1}$.
To reproduce the kinetics of $\sigma^F$-AB$\cdot$ADP binding we have
to assume that nucleotide-free AB can also interact weakly with
$\sigma^F$ ($k_{off} = 40$ s$^{-1}$). The $\sigma^F \cdot$AB
affinity in the presence of AA (77-93) was determined from
fluorescence quenching data on AA-induced $\sigma^F$ release.
Fitting of the experimental data required the offrate with ATP
present to be 100 times higher in the presence of  AA than in its
absence (a feature that had been noted previously
\cite{Clarkson04_2}), and required an even higher off-rate in the
presence of ADP. We set the off-rate in the presence of ADP to the
value of the on-rate, thus assuming no binding, but smaller values
for the off-rate would give similar simulation results.

\subsubsection*{Rates for AB conformational changes}
The SPR experiments on AB-AA binding provide the only rates
available for the interconversion between R and T-states, and we
used the rate for the disfavoured direction ($2$ s$^{-1}$)
throughout (116-137). In order to obtain  allosteric behaviour, the
favoured direction (94-115) when both monomers have symmetrical
ligands must be attained faster than the rate estimated by SPR in
the presence of ADP when only one monomer has bound AA (7.4
s$^{-1}$). We used $5 \times 10^3$ s$^{-1}$. The exact value for
this depends on the choice of the lid closure rate in the low
affinity conformation (see below).

\subsubsection*{ADP-ATP exchange}
The AB nucleotide binding pocket is partially covered by a flexible
loop known as the  ``ATP-lid'' \cite{Campbell02}. ADP-ATP exchange
(138-182) requires lid opening \cite{Clarkson04_1} and subsequent
release of ADP and binding of ATP. The affinity of AB for ATP and
ADP has been determined to be $K_D \sim 200$ $\mu$M \cite{Lord96}.
Fluorescence experiments suggest a slightly lower AB$\cdot$ADP
affinity \cite{Clarkson04_1}. Assuming that the on-rate for this
interaction is in the range found for other associations between
proteins and small ligands ($k_{on} \sim 10^7$ M$^{-1}$s$^{-1}$
\cite{Fersht}) this gives $k_{off} \sim 2 \times 10^3$ s$^{-1}$.
ADP-ATP exchange has been found experimentally to proceed at a rate
of about 1 s$^{-1}$. This is much slower than would be expected from
the nucleotide on- and off-rates, and has been interpreted as the
lid opening rate \cite{Clarkson04_1}. The nucleotide-bound monomer
is predominantly in the low-affinity conformation, and lid opening
in this conformation was therefore taken to proceed at 1 s$^{-1}$.
Given the choice for the rate of the conformational change, the lid
closure rate had to be set to $5\times 10^5$ s$^{-1}$ in order to
reproduce the observed rebinding rate of $\sigma^F \cdot$AB
complexes after the AA-induced $\sigma^F$ release (Fig.
\ref{Fig_par_det}B), in agreement with the intuitive consideration
that the lid should be predominately closed when nucleotides are
bound. It should be noted that these several steps associated with
ADP-ATP exchange were included in the model only when the level of
detail of available experimental information required a similar
level of detail in the model - otherwise only the rate-limiting
steps were considered in order to reduce the complexity of the
model.

\subsubsection*{ AA phosphorylation and dephosphorylation rates}
\begin{figure} [!b]
\begin{center}
\includegraphics[width=15cm]{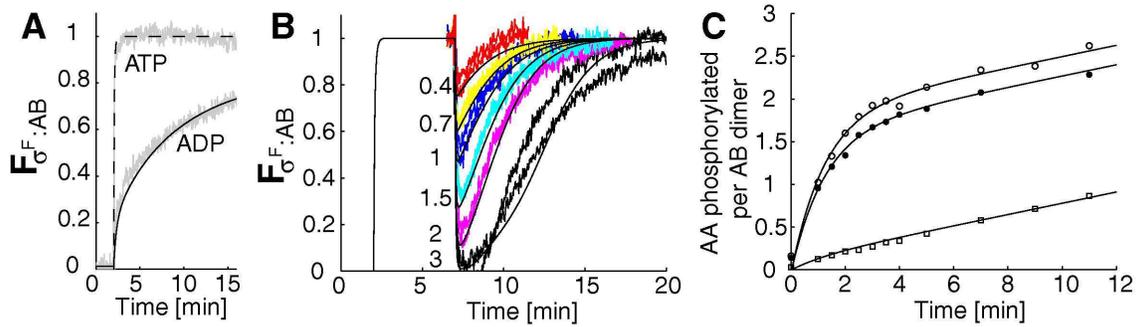} \caption
{\label{Fig_par_det} {\small \textbf{Estimate of parameter values.}
\textbf{(A)} Interactions between 1.3 $\mu$M $\sigma^F$ and 1 $\mu$M
AB. The results of fluorescence quenching experiments
\cite{Clarkson04_3} are shown as output from a recorder, and
simulations are shown as smooth lines, either dashed (for
ATP-dependent interactions) or continuous (for ADP-dependent
interactions). \textbf{(B)} $\sigma^F$ release from AB$\cdot$ATP
upon addition of AA. The results from two experiments are shown for
the disruption of $\sigma^F\cdot$AB complexes by 0.4 (red); 0.7
(yellow); 1 (blue); 1.5 (cyan); 2 (magenta); 3 (black) $\mu$M AA.
The corresponding simulations are shown in smooth black lines, which
are labeled with the relevant concentrations of AA. Experiments were
carried out as described in \cite{Clarkson04_3} with the fluorescent
mutant $\sigma^F$-W46L. \textbf{(C)} Phosphorylation of AA by AB
when AB is directly incubated with 40 $\mu$M AA and 100 $\mu$M ATP
($\circ$) or pre-incubated for 5 minutes with either 5 $\mu$M ADP
($\bullet$) or 5 $\mu$M ADP and 40 $\mu$M AA ($\square$). The
symbols show the experimental results \cite{Clarkson04_1},
normalised by the factor 1.3 to take account of inaccuracies in the
protein determination (see main text). The lines show the
predictions of the mathematical model.}}
\end{center}
\end{figure}

The phosphorylation of AA (183-192) exhibits a biphasic timecourse
(Fig. \ref{Fig_par_det} C) \cite{Najafi97,Magnin97}. The initial
rate is equal to the AA phosphorylation rate, and the subsequent
reduction in rate is due to the sequestering of AA and AB in
AB$\cdot$ADP$\cdot$AA complexes. The phosphorylation rate of AA can
thus be estimated from the initial slope of the biphasic
phosphorylation reaction as 0.013 s$^{-1}$, a value that is close to
previous estimates \cite{Shu04}. The rate of AA-P production when
AB$\cdot$ADP$\cdot$AA complexes have already formed is much higher
than the rate of dissociation of the AB$\cdot$ADP$\cdot$AA complex.
ADP-ATP exchange must therefore be possible while AA is bound to
AB$\cdot$ADP (162-165) - though at a reduced rate. Modelling of the
steady-state rate of phosphorylation requires a ADP-ATP exchange
rate for AB$\cdot$AA of $1.2 \times 10^{-3}$ s$^{-1}$ (Fig.
\ref{Fig_par_det}C). ATP hydrolysis will be much slower than AA
phosphorylation and was therefore ignored.

Interestingly, when AB is pre-incubated for 5 minutes with AA and
5$\mu$M ADP before addition of 100 $\mu$M ATP the biphasic
phosphorylation timecourse can be directly transformed into a linear
timecourse with the lower phosphorylation rate throughout (Fig.
\ref{Fig_par_det} C); by contrast, pre-incubation with 5$\mu$M ADP
alone has only a minor effect (Fig. \ref{Fig_par_det} C). From these
results it can be concluded that ADP-AB binding is weak but can be
facilitated by AA, which itself binds weakly to AB when no
nucleotide is bound \cite{Clarkson01}. In order to obtain such weak
AB-ADP binding (while retaining a high affinity once ADP is bound),
the R state must be favoured when the lid is open; in this
conformation AB must have a lower affinity for ADP ($k_{off} = 10^5$
s$^{-1}$) and the lid must be less mobile, that is the lid opening
and closure rates need to be of order $10^{-4}$ s$^{-1}$. The only
exception to this is when $\sigma^F$ is bound, when we have to
assume a lid closure rate of 70 s$^{-1}$ to reproduce $\sigma^F
\cdot$AB.ADP binding (Fig. \ref{Fig_par_det}A). In order to limit
the number of variables we ignored the (disfavoured) closed lid
state of AB when no nucleotide is bound; instead we assumed that the
R state was 50 times more abundant than we had assumed elsewhere in
the model. This is important in order to reduce the fraction of AB
in the open, nucleotide-free tense state, which has a higher
affinity for ADP than the relaxed state. Note that these particular
parameter choices are needed to reproduce the \textit{in vitro}
results shown in Figure \ref{Fig_par_det} but have
little or no impact on the other results shown.\\

Dephosphorylation of AA-P (193-195) \textit{in vitro} with IIE
domain III has been measured to proceed at 0.085 s$^{-1}$
\cite{Lucet00}. The on-rate for IIE$\cdot$AA-P is unknown, as is the
off-rate, and these were set to ``standard'' protein binding and
unbinding rates, e.g. $10^6$ M$^{-1}$s$^{-1}$ and 0.1 s$^{-1}$
respectively.

\section*{Verification of the model by comparison to \textit{in vitro} data}
\subsection*{The model quantitatively reproduces experimental
results of $\sigma^F \cdot$AB sensitivity to variations in AA and
IIE.} 
\begin{figure} [!b]
\begin{center}
\includegraphics[width=15cm]{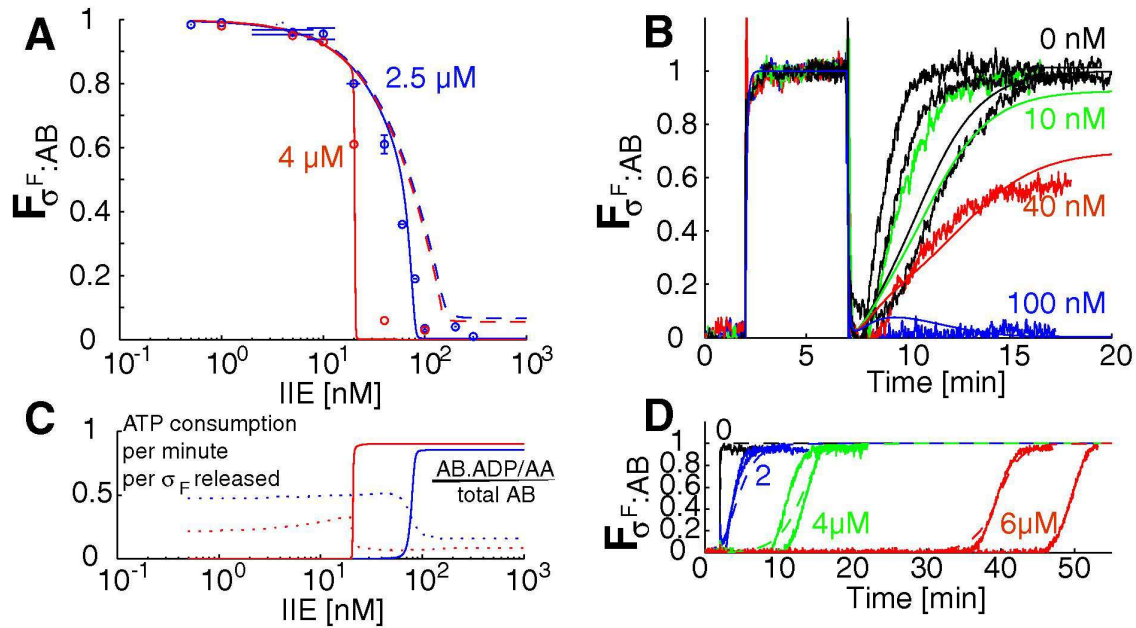}
\caption {\label{Fig_result_invitro} {\small \textbf{The
dissociation of $\sigma^F\cdot$AB complexes is sensitive to changes
in the concentration of IIE and AA. } \textbf{(A)} Steady state
after IIE-dependent dissociation of $\sigma^F\cdot$AB$\cdot$ATP
(1.3$\mu$M $\sigma^F$, 1$\mu$M AB dimer) in the presence of
2.5$\mu$M (blue) or 4$\mu$M (red) AA. Circles represent experimental
results \cite{Clarkson04_3}. Continuous lines show the predictions
of the mathematical model, which includes the formation of the
AB$\cdot$ADP$\cdot$AA complex. Dashed lines show simulations based
on the assumption that the formation of such a complex is
impossible. \textbf{(B)} Kinetics of IIE-dependent dissociation of
$\sigma^F\cdot$AB$\cdot$ATP (1.3$\mu$M $\sigma^F$, 1$\mu$M AB dimer)
in the presence of 2.5$\mu$M AA, in the absence of IIE (black lines)
or with 10 (green), 40 (red) or 100 (blue) nM IIE. Experimental
results are taken from \cite{Clarkson04_3}, and are shown as output
from a recorder. Note that three experiments were run in the absence
of IIE. The smooth continuous lines show simulations from the
mathematical model. \textbf{(C)} Simulations of the proportion of AB
sequestered into AB$\cdot$ADP$\cdot$AA complexes (continuous line)
and ATP consumption per minute per $\sigma^F$ released (dotted line)
in the presence of 2.5$ \mu$M (blue) or 4$ \mu$M (red) AA.
\textbf{(D)} $\sigma^F \cdot$AB complex formation on mixing of 1
$\mu$M AB and 1.3 $\mu$M $\sigma^F$ with no (black), 2 $\mu$M
(blue), 4 $\mu$M (green), or 6 $\mu$M (red) AA. Experimental results
(two for each concentration of AA) are taken from
\cite{Clarkson04_3} and are shown as output from a recorder; the
dashed lines show simulations from the mathematical model at the
same concentrations of AA.}}
\end{center}
\end{figure}

With all the kinetic parameters determined we first sought to
reproduce recent biochemical \textit{in vitro} results (shown as
circles in Fig. \ref{Fig_result_invitro} A) which revealed that
$\sigma^F \cdot$AB$\cdot$ATP complexes are highly sensitive to
changes in IIE activity \cite{Clarkson04_3}, thus lending support to
the hypothesis that the septation-dependent increase in IIE activity
may lead to $\sigma^F$ release in the prespore \cite{Duncan95}.
Translation of the scheme into a set of coupled differential
equations allowed us to reproduce quantitatively the high
sensitivity of $\sigma^F \cdot$AB$\cdot$ATP complexes to such
changes in IIE activity, and also showed that the complexes are
sensitive to changes in the AA concentration (compare 2.5 $\mu$M and
4 $\mu$M curves in Fig. \ref{Fig_result_invitro} A). The model also
captured the kinetics of $\sigma^F \cdot$AB complex formation upon
addition of ATP, the dissociation of the complex upon addition of AA
and the effect of IIE on re-formation of the complex (Fig. 2C in the
main paper).  The considerable variation in the experimental
timecourses (compare the three black curves for IIE=0 nM in Fig. 2C
in the main paper) may be due to the high sensitivity of the
kinetics to concentration. Comparison to the results in Fig.
\ref{Fig_par_det}B suggests that the slower recovery rate found in
the simulation is the more realistic one. The values for the
steady-state dissociation of the complex are more robust, and the
standard deviations between three independent experiments, as
indicated by error bars in Figure \ref{Fig_result_invitro}A, are
generally small. In both cases the modeling results are within the
experimental range. The close fit of simulation and experiment,
despite inherent experimental errors in the parameters including
protein concentration, can be attributed to the robustness of the
model (see below) and to the fact that the constants for the various
on- and off-rates were all derived from similar experimental
protocols.

\subsection*{The robustness of the mathematical model to variations in parameters.}
The regulatory network can be regarded as robust if small changes in
the parameter values lead to only small changes in the response.
Given the availability of experimental data, the sensitivity of the
model to variations in the kinetic parameters can be assessed by
determining the deviation $\Delta_j$ of the simulation values $s_i$
from the $N$ experimental equilibrium results $e_i$ (Fig.
\ref{Fig_result_invitro}A) in response to a change in the kinetic
parameter value $k_j$ as

\begin{equation}
\Delta_j =
\sqrt{\frac{1}{N}\sum_{i=1}^{N}\left(\frac{s_i-e_i}{e_i}\right)^2}.
\end{equation}

Those reactions $j$, for which $\Delta_j$ did not increase by more
than $1\%$ when $k_j$ was either set to zero or varied between
$10^{-3}$ and $10^6$ in multiples of 10 were considered insensitive.
Those for which even a small (10-fold) change in the parameter led
to an increase in $\Delta_j$ by more than $1\%$ were considered
sensitive. This analysis picks out nicely the key reaction pathways
in the scheme and shows that relatively few reactions are sensitive
to parameter variations. Sensitive reactions include the
phosphorylation and dephosphorylation of AA, the binding of AA to
$\sigma^F \cdot$AB complexes and also those that lead to
AB$\cdot$ADP$\cdot$AA complex formation or ADP-ATP exchange.\\

\subsection*{Predictions of the effect of AB$\cdot$ADP$\cdot$AA complex formation.}
AB$\cdot$ADP$\cdot$AA complex formation coincides with $\sigma^F$
release from the $\sigma^F\cdot$AB$\cdot$ATP complex (compare Fig.
\ref{Fig_result_invitro} A and B) and is important for the high
sensitivity of $\sigma^F$ release to changes in the AA
concentration. The formation of AB$\cdot$ADP$\cdot$AA complexes also
minimizes the need for cycling between AA and AA-P \cite{Magnin97}
and associated ATP consumption. Thus, the ATP consumption per minute
per $\sigma^F$ released does not increase monotonically with
increasing IIE concentrations, but reaches a minimum (Fig.
\ref{Fig_result_invitro} B, dotted  line) when sufficient unbound AA
is available to sequester AB in AB$\cdot$ADP$\cdot$AA complexes
(Fig. \ref{Fig_result_invitro} B, solid line).

The AA-dependent trapping of AB$\cdot$ADP in AB$\cdot$ADP$\cdot$AA
complexes can be observed experimentally in the form of an
increasing delay in $\sigma^F \cdot$AB complex re-formation when
equimolar concentrations of AB and $\sigma^F$ are mixed with
increasing molar ratios of AA (Fig. 2D in the main paper). The
non-linear increase in the delay times with increasing AA
concentrations reflects the non-linear AA-dependence of
AB$\cdot$ADP$\cdot$AA complex formation. It should be noted that the
extent to which this delay increases with increasing AA
concentration would be smaller if the AA-AB$\cdot$ADP affinity were
lower.

\section*{Application of the model to the sporulating cell \textit{in vivo} }
\begin{figure} [!b]
\begin{center}
\includegraphics[width=15cm]{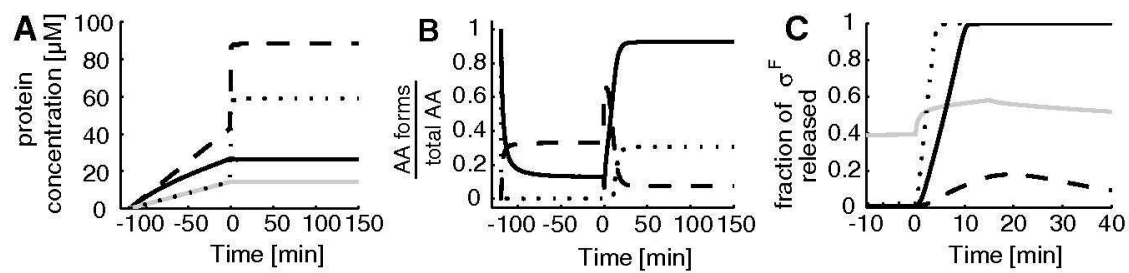} \caption
{\label{Fig_phys_prot_express} {\small \textbf{The simulation
reproduces the experimentally observed concentrations of sporulation
proteins.} \textbf{(A)} Increase in the concentration of sporulation
proteins during sporulation (AA- dashed; AB - continuous black; IIE
- dotted; $\sigma^F$ - grey).  \textbf{(B)} Time-dependent fraction
of AA that is unphosphorylated (solid line), unbound and
unphosphorylated (dotted line) or bound in IIE$\cdot$AA-P complexes
(dashed line). In both panels, sporulation-dependent protein
expression is started at -120 minutes; a four-fold increase in IIE
and associated AA-P (to simulate the effect of asymmetric septation)
is introduced at t=0. \textbf{(C)} Fraction of $\sigma^F$ released
in simulations based on the assumption that the IIE activity
\textit{in vivo} was not inhibited (continuous grey line), or
inhibited by 10 (dotted), 20 (continuous black) or 40 times (dashes)
compared with the maximum IIE activity measured \textit{in vitro}. A
four-fold increase in IIE and associated AA-P (to simulate the
effect of asymmetric septation) is introduced at t=0.}}
\end{center}
\end{figure}

With a model that reproduces the \textit{in vitro} data
quantitatively we are now in a position to investigate the
physiological situation. Relatively few alterations have to be made
to the model to take account of the differences between the
\textit{in vitro} and \textit{in vivo} conditions. The membrane
location of IIE in the cell does not need to be considered
explicitly in the model since, given the dimensions of  the
forespore, spatial concentration gradients on the timescale of the
reactions would be expected to be negligible. A change has, however,
to be made to the ATP and protein concentrations, which are higher
\textit{in vivo} at the time of septation than in the \textit{in
vitro} experiments. The ATP concentration was set to 1 mM because
the lower end of the experimentally reported ATP concentration range
(0.85 mM to 3 mM) for vegetative \textit{B. subtilis}
\cite{Jolliffe81,Guffanti87,Hecker88}) is likely to apply to the
starving sporulating condition. The ADP concentration was set to 100
$\mu$M, which is a tenth of the ATP concentration. To simulate the
experimentally observed time course of protein expression during
sporulation \cite{Magnin97,Lord99,Lucet99} we started the
sporulation reaction at time -120 minutes without protein and used
an expression rate of $6 \times 10^{-9}$ M s$^{-1}$ for AA and AB,
whose translation has been suggested to be coupled \cite{Fort84},
and $2 \times 10^{-9}$ M s$^{-1}$ for $\sigma^F$ and IIE (Fig.
\ref{Fig_phys_prot_express}A). Septation (a 4-fold increase in IIE
and associated AA-P) was introduced after 120 minutes
\cite{Magnin97,Lord99,Lucet99} unless otherwise stated. The
transient genetic imbalance, which arises upon septum formation and
which results in the prespore lacking the \textit{spoIIA} operon for
10-20 minutes after septation \cite{Frandsen99,Dworkin01}, was
modelled by prohibiting expression of AA, AB and $\sigma^F$ for 15
minutes after septation. The emergence of unphosphorylated AA (Fig.
\ref{Fig_phys_prot_express}B) further leads to repression of
\textit{spoIIE} and the \textit{spoIIA} operon via inactivation of
Spo0A \cite{Arabolaza03}. Note that in agreement with experimental
observations \cite{King99} almost half of all unbound AA (e.g. AA
that is not bound to IIE) is unphosphorylated already before
septation (Fig. \ref{Fig_phys_prot_express}B).

An amino-acid sequence in the extreme C-terminus of AB markedly
increases the AB degradation rate \textit{in vivo} \cite{Pan01}. In
the simulations unbound AB was therefore associated with a
degradation rate of $4.1 \times 10^{-4}$ s$^{-1}$, which gives rise
to the experimentally observed half-life of about 28 minutes
\cite{Pan01}.

The only rate constant that needs to be altered from the \textit{in
vitro} model is the IIE phosphatase rate, which experiments suggest
is lower \textit{in vivo} \cite{Feucht02}. Modelling predicts that
the reduction needs to be by at least 10-fold to avoid
septation-independent release of $\sigma^F$ (Fig. 3A in the main
paper); a 20-fold reduction is used in all simulation results
presented here. This reduction can be partly accounted for by the
different \textit{in vivo} and \textit{in vitro} concentrations of
Mn$^{2+}$ and Mg$^{2+}$. The phosphatase rate is substantially
higher in the presence of Mn$^{2+}$ \cite{Schroeter99}; but, given
that the phosphatase rate has been determined in the presence of
Mn$^{2+}$ while Mg$^{2+}$ predominates in the bacterial cell, the
rate \textit{in vivo} may be considerably smaller than the
literature value. However, this is unlikely to account for a 10-fold
(let alone a 20-fold) reduction, and it may be that further
alterations must be made to the IIE environment or that an unknown
IIE inhibitor exists \cite{King99,Feucht02}.

\section*{IIE accumulation on the septum, AB allostery and a short half-life enable
compartment-specific gene transcription} While septal accumulation
of IIE accumulation \cite{Duncan95} and a short half-life of unbound
AB \cite{Pan01} have been established experimentally as key
components of the mechanism that enables prespore-specific
$\sigma^F$ release, it is unclear how they contribute and whether
they are sufficient given that the IIE activity increase in the
prespore is small and that the half-life of unbound AB strongly
exceeds the time scale on which $\sigma^F$ is activated. Moreover,
it has so far been a paradox, how released $\sigma^F$ can displace
$\sigma^A$ on the RNA polymerase despite the (unfavourable)
competition of $\sigma^F$ with the transcription factor $\sigma^A$
\cite{Lord99}.

We will show in the following that septal accumulation of IIE is
sufficient because IIE accumulates on the septum together with its
substrate AA-P; this increases the rate of AA formation by at least
2.5-fold. The here-described allostery of AB enables the high
sensitivity to such small changes in AA production; degradation of
unbound AB is important to tune the protein concentrations in a way
that small changes can act as a trigger without rendering the
regulatory network vulnerable to fluctuations in protein expression.
The relative affinities of the RNA polymerase for $\sigma^F$ and
$\sigma^A$ are shown to be optimized to allow for a coexistence of
both holoenzymes. Other hypotheses such as the removal of a IIE
inhibitor or effects of the transient genetic imbalance are shown to
be irrelevant.

\subsection*{Accumulation of IIE (in association with its substrate AA-P)
at the septum is sufficient to induce $\sigma^F$ release in the
prespore.} 
\begin{figure} [!b]
\begin{center}
\includegraphics[width=15cm]{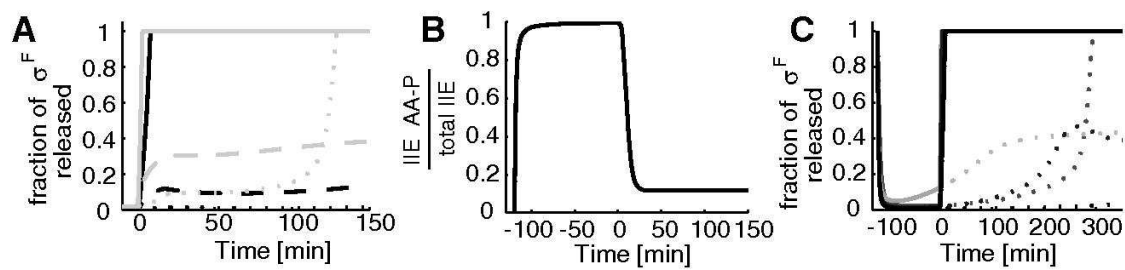} \caption
{\label{Fig_phys_EAp} {\small \textbf{Simulations showing that a
septation-dependent increase in IIE$\cdot$AA-P is sufficient for
$\sigma^F$ release.} \textbf{(A)}  $\sigma^F$ release in response to
a 4-fold increase in IIE alone (dashes) in AA-P alone (dotted), or
in IIE and AA-P (continuous line) imposed at t=0. The IIE
phosphatase rate was either 10-times (grey) or 20-times (black)
smaller than \textit{in vitro}. \textbf{(B)} Time-dependent fraction
of IIE in IIE$\cdot$AA-P complexes. \textbf{(C)} Time-dependent
$\sigma^F$ release in wildtype (black), when AB expression is
reduced by $30\%$ (red line), if IIE expression is increased by
two-fold (green and blue lines) and if IIE activity is increased by
two-fold (green line). In all panels, the septum was assumed to form
at t=0 (see main text). }}
\end{center}
\end{figure}

Contrary to previous suggestions, the model demonstrates
that septal accumulation of IIE is sufficient to induce $\sigma^F$
release in the prespore; it is not necessary to increase the IIE
phosphatase rate in the prespore (for example by the removal of an
inhibitor). As AA production is equal to the phosphatase rate times
the concentration of IIE$\cdot$AA-P complexes, it is sufficient to
increase the substrate and enzyme concentrations by accumulation on
the septum (Fig. \ref{Fig_phys_EAp}A). Substrate and enzyme increase
together when IIE accumulates on the septum because, owing to low
phosphatase rate and the slightly higher total AA concentration
relative to the IIE concentration in the cell, almost all IIE is
associated with AA-P before septation (Fig. 3B in the main paper);
enzyme and substrate would accumulate together on the septum even if
the IIE$\cdot$AA-P affinity were as low as micromolar.

An increase in either the IIE activity or AA-P concentration alone
is far less efficient even if the phosphatase rate is set to the
highest physiological possible value (Fig. \ref{Fig_phys_EAp} A). It
is therefore necessary for IIE to be present before septation, since
that is the only way in which septal accumulation of IIE could also
increase the AA concentration in the prespore. In fact asymmetric
septation appears to require the presence of a significant IIE
concentration \cite{Barak96,Khvorova98,Feucht99,Ben-Yehuda02}.

%While the release of an inhibitor is not necessary to drive
%$\sigma^F$ release, available experimental data suggest that the
%reduction of the \textit{in vivo} phosphatase rate is at least in
%part dependent on a titratable inhibitor. Thus $\sigma^F$ release is
%found to be delayed by only $30\%$ in mutants that overexpress IIE
%but cannot form a septum \cite{Arigoni99}. The simulation agrees
%with the interpretation that IIE activity needs to be higher than in
%the wildtype experiments to capture this result (Fig.
%\ref{Fig_phys_EAp} B, compare green and blue curves). Without such
%activity increase $\sigma^F$ activation should be similar in mutants
%that overexpress IIE or express less AB (Fig. \ref{Fig_phys_EAp} B,
%compare blue and red curves). The model further predicts that the
%(so far unnoticed \cite{Arigoni99}) effects of a 30-fold reduction
%in AB expression on $\sigma^F$ activation should become apparent if
%the experiments were repeated with mutants that cannot form a
%septum.

\subsection*{Formation of the $\sigma^F$-RNA polymerase holoenzyme }
Prespore-specific $\sigma^F$ release is only a first step to
differential gene expression. In order to direct gene expression
$\sigma^F$ needs to bind to the RNA polymerase, from where it needs
to displace $\sigma^A$. The mechanism of the latter has so far been
unclear given the (unfavourable) competition of $\sigma^F$ with
$\sigma^A$ ($\sigma^A$ binds to the polymerase with a 25-fold higher
affinity (560 nM and 22.5 nM for $\sigma^F$ and $\sigma^A$
respectively), but it is only slightly less abundant ($\sim 7.5 \mu$
M) than released $\sigma^F$ \cite{Lord99}). As the core RNA
polymerase concentration is comparatively high($\sim 7.5 \mu$ M),
the mathematical model predicts that micromolar concentrations of
$\sigma^F \cdot$RNApol holoenzyme can form spontaneously in response
to asymmetric septation (Fig. 3E in the main paper), and that these
complexes co-exist with micromolar concentrations of $\sigma^A
\cdot$RNApol holoenzyme (Fig. 3E in the main paper). The latter
result is in agreement with the experimental observation that
$\sigma^A \cdot$RNApol holoenzyme persists throughout the
sporulation process \cite{Fujita98}.

Figure 3E in the main paper further shows that IIE (and AA-P bound
to it) needs to increase by at least 2.5-fold in order to ensure the
formation of micromolar $\sigma^F \cdot$RNApol concentrations. Given
the difference in volume between the two compartments, accumulation
of IIE on the septum will in fact lead to at least such an increase
(see Supplementary Information). However, as discussed in the
Supplementary Information, a 4-fold increase is a more realistic
description of the sporulating cell and is therefore used in all the
simulations presented here.

\subsection*{The role of the transient genetic imbalance}
The view that $\sigma^F$ release is driven by a change in IIE
activity at the asymmetric septum has been challenged by the finding
that $\sigma^F$ was still released, albeit with a significant delay,
in mutants where wild-type IIE was replaced by a soluble phosphatase
domain (IIE$\Delta$mem) \cite{Arigoni99}. The formation of $\sigma^F
\cdot$RNApol in the IIE$\Delta$mem mutant was subsequently shown to
be enabled by the block in AB expression that follows septation due
to a transient genetic imbalance \cite{Dworkin01}.

\begin{figure} [!b]
\begin{center}
\includegraphics[width=10cm]{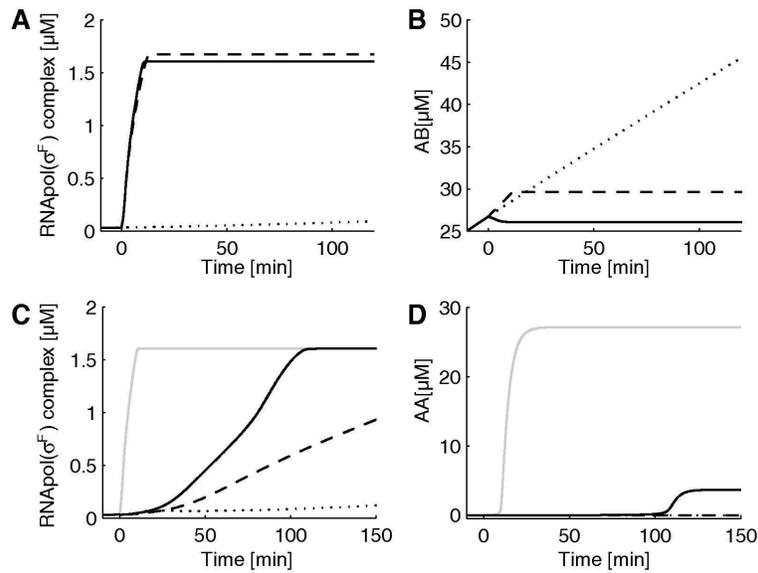}
\caption {\label{Fig_phys_IIE_altern_mod} {\small \textbf{The
 effect of a transient genetic imbalance on $\sigma^F$ release.}
\textbf{(A)} Formation of RNApol($\sigma^F$) complexes, \textbf{(B)}
total AB concentration for wt (continuous line)
\textit{spoIIAB}(ori) (dashes) IIE$\Delta$mem $\&$
\textit{spoIIAB}(ori) (dotted). \textbf{(C)} Formation of
RNApol($\sigma^F$) complexes, \textbf{(D)} total AA concentration
for wt (continuous grey line), for IIE$\Delta$mem if chromomosome
translocation is blocked (solid line) if \textit{Spo0A} is
constituively repressed (dashed line) or if IIE$\Delta$mem $\&$
\textit{spoIIAB}(ori) (dotted). In all panels, the septum was
assumed to form at t=0 (see main text).}}
\end{center}
\end{figure}

The simulation correctly predicts the phenotype of mutants that lack
the transient genetic imbalance and the impact of the IIE$\Delta$mem
on such background (Fig. \ref{Fig_phys_IIE_altern_mod}A).
Importantly however, the simulation also shows that a 10-20 minute
block in AB expression (as may arise from the transient genetic
imbalance - Fig. \ref{Fig_phys_IIE_altern_mod}B) is insufficient to
drive prespore specific $\sigma^F$ release in IIE$\Delta$mem mutants
(Fig. \ref{Fig_phys_IIE_altern_mod}C, compare solid and dotted
lines). A continuous block in AB expression is necessary (Fig.
\ref{Fig_phys_IIE_altern_mod}C, solid line) and this cannot be
achieved by Spo0A repression since, unlike wildtype, IIE$\Delta$mem
does not dephosphorylate sufficient AA upon septation (Fig.
\ref{Fig_phys_IIE_altern_mod}D).

%A possible explanation for continued repression of protein
%expression is that the failure to localize IIE to the membrane
%impairs the translocation of the bacterial chromosome such that the
%prespore lacks \textit{spoIIA} for a longer time. Two experimental
%observations are supportive of this hypothesis. First of all, a
%mutant that co-expresses a phosphatase deficient version of IIE on
%the membrane with the soluble phosphatase domain leads to a smaller
%sporulation efficiency \cite{Arigoni99}. Secondly, the
%experimentally observed kinetics resemble rather the kinetics
%predicted for repression of \textit{spoIIA} and expression of
%\textit{spoIIE} (Fig. \ref{Fig_phys_IIE_altern_mod}C, compare solid
%and dashed lines) while \textit{spoIIE} would also be repressed upon
%repression of Spo0A. Interestingly, about half of all IIE$\Delta$mem
%mutants release $\sigma^F$ in the entire cell despite of a slightly
%lower expression of the truncated protein \cite{Arigoni99}. This is
%likely to be the consequence of a higher IIE phosphatase rate for
%the smaller protein, as would be in agreement with experimental
%observations which suggest that the phosphatase rate is regulated by
%a domain close to the membrane \cite{Feucht02}. The removal of the
%membrane anchor may then relief relief inhibition with the effect
%that a considerable fraction of cells achieves a critical IIE
%activity already before septation or afterwards in both
%compartments.

We conclude that a genetic imbalance as short as 10-20 minutes
cannot drive compartment specific gene expression. AB decay leads to
$\sigma^F$ release only after 30-40 minutes and therefore only
enables significantly delayed $\sigma^F$ release in IIE$\Delta$mem
mutants. Successful sporulation, however, requires that  $\sigma^F$
is released within 10 minutes after septation \cite{Hilbert04}. IIE
accumulation on the septum is therefore necessary to enable rapid
$\sigma^F$ release. Additional mechanisms such as inhibitor removal
\cite{Frandsen99} or a block in the release of dephosphorylated AA
from IIE before septum formation \cite{King99} are not required.

\subsection*{The role of the short half-life for unbound AB}
While results like those above question a role for a short half-life
of unbound AB in rapid prespore-specific $\sigma^F$ release, mutants
with stable AB are sporulation deficient \cite{Pan01}. The
simulation now reveals that the 28 minutes half-life of unbound AB
is important to limit the AB concentration before septation (Fig. 3D
in the main paper). The use of AB degradation rather than a lower
expression rate is important to guarentee robustness to stochastic
fluctuations in gene expression as is discussed below.

\subsection*{Allosteric behaviour of AB ensures both sensitivity and robustness}
\begin{figure} [!b]
\begin{center}
\includegraphics[width=10cm]{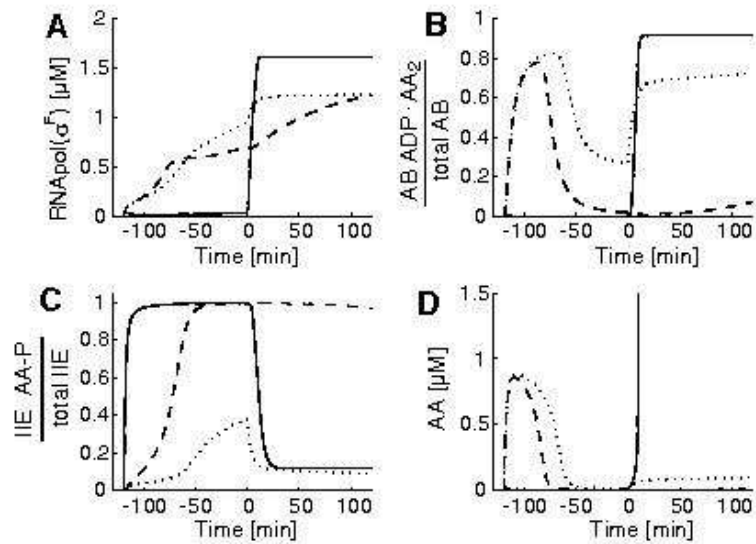}
\caption {\label{Fig_allo} {\small \textbf{The allosteric behaviour
of AB is necessary for successful sporulation.} \textbf{(A)} The
formation of RNApol($\sigma^F$) holoenzyme, \textbf{(B)} the
fraction of AB in AB$\cdot$ADP$\cdot$AA complexes, \textbf{(C)} the
fraction of IIE in IIE$\cdot$AA-P complexes, and \textbf{(D)} the
fraction of unphosphorylated AA if AB is allosteric (continuous
line) or not allosteric and the activity of IIE is inhibited 20-fold
(dotted) or 1000-fold (dashes) compared with the maximum activity
measured \textit{in vitro}. A four-fold increase in IIE, together
with its associated AA-P, is imposed at t= 0. }}
\end{center}
\end{figure}

AB allostery is key to the high sensitive to changes in the
IIE$\cdot$AA-P concentration and ensures that the system is robust
to the initial parallel increases in the protein concentrations
(Fig. \ref{Fig_allo}A). Due to AB allostery $\sigma^F \cdot$AB
complexes are stable at low AA concentration but become rapidly
dissolved once the AA concentration reaches a critical threshold
after septation. The premature formation of AB$\cdot$AA complexes in
the absence of allostery (Fig. \ref{Fig_allo}B) would have two
detrimental effects. First, sequestration of AB would free
$\sigma^F$ (Fig. \ref{Fig_allo}A,B). Secondly, sequestration of AA
would reduce the fraction of IIE that is bound by its substrate AA-P
(Fig. \ref{Fig_allo}C), thereby interfering with the simultaneous
accumulation of IIE and AA-P when the septum forms, and thus
limiting the formation of unphosphorylated AA in the prespore since
- unlike in metabolic pathways - IIE's substrate (AA-P) is a protein
and present at a low concentration in the cell.

We have investigated whether non-allosteric behaviour of AB could be
compensated for by a stronger inhibition of IIE. However, the model
shows that even a 1000-fold inhibition of IIE activity would be
ineffective in preventing premature $\sigma^F$ release (Fig
\ref{Fig_allo}, dashed lines) and would fail to provide for specific
$\sigma^F$ release in response to septation.

\section*{Economic efficiency through limiting ATP consumption} 
\begin{figure} [!b]
\begin{center}
\includegraphics[width=8.5cm]{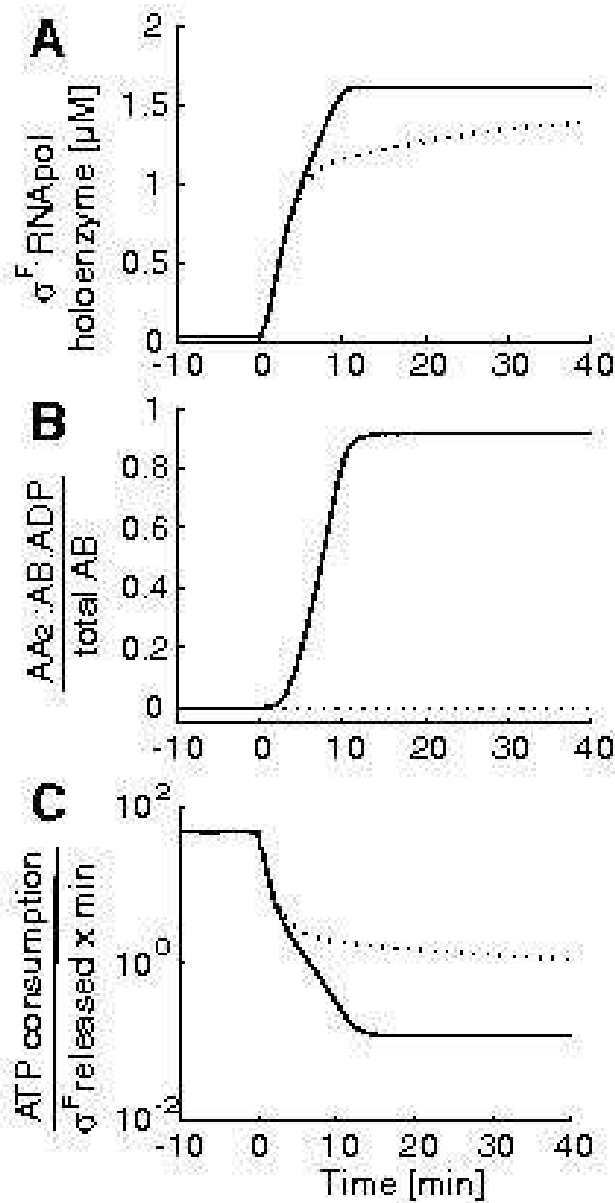} \caption
{\label{Fig_phys_AB_ADP} {\small \textbf{AB$\cdot$ADP$\cdot$AA
complexes form upon asymmetric septation.} The simulations show the
effect of a four-fold increase at t=0 min of IIE, together with its
associated AA-P, on \textbf{(A)} formation of RNApol($\sigma^F$)
holoenzyme, \textbf{(B)} fraction of AB in AA$_2\cdot$AB$\cdot$ADP
complexes. \textbf{(C)} ATP consumption per minute per  $\sigma^F$
released for wildtype (continuous line) or for a variant in which
AB$\cdot$ADP$\cdot$AA cannot be formed (dotted).}}
\end{center}
\end{figure}

The
avoidance of ``ATP wastage'' can be expected to become particularly
important once nutrients are scarce as is the case at the onset of
sporulation. The regulatory network that controls $\sigma^F$ release
is optimized to limit ATP wastage. Rapid formation of $\sigma^F
\cdot$RNApol holoenzyme (Fig. \ref{Fig_phys_AB_ADP}A) is accompanied
by the accumulation of most of the AB in AB$\cdot$ADP$\cdot$AA
complexes (Fig. \ref{Fig_phys_AB_ADP}B). If the simulation is
altered so as to prohibit the formation of these complexes, AB
accumulates instead in AB$\cdot$ATP$\cdot$AA complexes; but the
consumption of ATP is then greatly increased (Fig. 4D in the main
paper). We conclude that, for $\sigma^F \cdot$RNApol holoenzyme to
be formed and to persist without excessive waste of ATP, the
formation of AB$\cdot$ADP$\cdot$AA complexes and the degradation of
AB are both important. AB$\cdot$ADP$\cdot$AA complexes therefore
play an important role in acting as a sink for AB after the increase
in IIE (and associated AA-P) that occurs at asymmetric septation.

\section*{Robustness}
\begin{figure} [!b]
\begin{center}
\includegraphics[width=8.5cm]{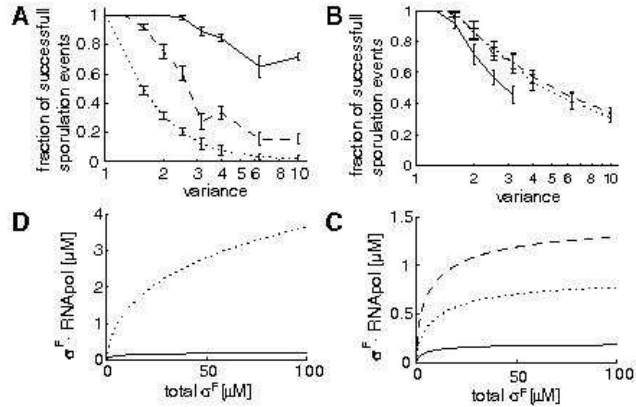}
\caption {\label{Fig_robust} {\small \textbf{The robustness of the
regulatory network.} ) \textbf{(A)} The fraction of successful
sporulation events if the rates of protein expression are varied
randomly by the factor given on the x-axis and all expression rates
are coupled (-), AA and AB expression is coupled ($\cdots$), all
expression rates are varied randomly, the AB expression rate is
$4\times10^{-9}$ M s$^{-1}$ and there is no AB degradation (- -) . A
successful sporulation event was defined as one for which the
concentration of RNA polymerase bound to  F is lower than 0.4 $\mu$M
before, and higher than 1 $\mu$M after, septation. Standard
deviations are based on 10 sets of 100 runs each. \textbf{(B)}
Fraction of successful sporulation events if either all binding
rates (dashed), all enzymatic rates (dotted), all binding and
enzymatic rates (solid) are varied randomly by the factor given on
the x-axis. Standard deviations are based on 10 sets of 100 runs
each; successful sporulation events are defined as those with a
concentration of RNA polymerase($\sigma^F$) holoenzyme less than 0.4
$\mu$M without, and greater than 1 $\mu$M with, a 4-fold increase in
IIE (and associated AA-P). \textbf{(C)} Concentration of RNA
polymerase($\sigma^F$) holoenzyme as a function of the reference
protein concentrations (as exemplified by the $\sigma^F$
concentration). The relative concentration of IIE (and the
associated AA-P) was either increased by four-fold (continuous line)
or left unchanged (dotted). \textbf{(D)} Concentration of RNA
polymerase $\sigma^F$ holoenzyme if the affinities of the core RNA
polymerase for $\sigma^F$ and $\sigma^A$ were as in the wild type
(continuous line) or made equal by varying the rate of dissociation
of either the $\sigma^F$ holoenzyme ($k_{off}(\sigma^F) = 0.02$
s$^{-1}$ - dashes) or the $\sigma^A$ holoenzyme ($k_{off}(\sigma^A)
= 0.55$ s$^{-1}$ - dotted).}}
\end{center}
\end{figure}

Biological networks need to be organized in a way that they are
sensitive to regulatory input but robust to the small variations
that may arise from changes in environmental conditions or from
stochastic fluctuations within the cell. The sporulation network
meets both conditions to an impressive degree. The protein
concentrations are tuned to ensure a high sensitivity to changes in
the IIE concentration while being robust to stochastic fluctuations
in gene expression rates (Fig. 4A in the main paper). The additional
robustness of the network to random changes in values in kinetic
parameter and enzymatic rates (Fig. \ref{Fig_robust}) justifies our
approach of using the kinetic parameters that were determined from
\textit{in vitro} experiments. By means of control coefficients we
determine the reactions that the regulatory network is most
sensitive to as the phosphorylation and phosphatase steps.

\subsection*{Protein expression is tuned to ensure robustness}
During the vegetative phase, the concentrations of  $\sigma^F$, AA,
AB, and IIE are too low to allow sufficient $\sigma^F \cdot$RNApol
formation to drive prespore development (Fig.
\ref{Fig_phys_prot_express}A and Fig. 4B in the main paper). This
efficiently protects the cell from premature $\sigma^F$ activation.
The triggering of differential gene expression requires the
formation of IIE$\cdot$AA-P complexes, and therefore the expression
of AA and IIE, prior to septation (see above). In order to avoid
premature $\sigma^F$ activation all sporulation proteins
($\sigma^F$, AA, AB and IIE) need to be expressed in parallel prior
to septation. Given the low numbers of proteins expressed, the
process is inherently stochastic. The ratio between the proteins is
essential to avoid spontaneous $\sigma^F$ release, and translational
coupling of all proteins is the best strategy to avoid failure
(dashed line in Fig. 4A in the main paper). However, the downside of
such coupling is that the ratios between proteins would then need to
be established by degradation which is uneconomic. The simulation
reveals that already the coupling of AA and AB translation strongly
increases robustness as compared to a mechanism that does not
involve coupling (compare dashed and dotted lines in Fig. 4A in the
main paper).

The robustness will even be higher than suggested by Figure 4A in
the main paper if the time point of septation depended on the
protein (and in particular the IIE) concentration as may indeed
apply to the cell \cite{Khvorova98,Ben-Yehuda02}. The impact of the
timepoint on sporulation efficiency reflects the need to exceed a
certain protein concentration before the septation-dependent
increase in IIE$\cdot$AA-P can trigger the formation of $\sigma^F
\cdot$RNApol at micromolar concentrations. The dotted line in Figure
4B in the main paper suggests this threshold to be 10 $\mu$ M for
$\sigma^F$ (and corresponding concentrations for the other proteins)
at the timepoint of septation which is within the physiological
concentration range for which $\sigma^F \cdot$RNApol formation has
been observed \cite{Magnin97,Lucet99}. We suggest that in order to
avoid unproductive septation, the timing of septum formation must be
linked to the overall concentration of the proteins relevant to
sporulation; this may explain the large variance in the delay
between the onset of sporulation and septation that is observed
under different sporulation conditions. In the absence of a
septation-dependent increase in the IIE$\cdot$AA-P concentration the
concentration of $\sigma^F \cdot$RNApol remains too low to trigger
downstream signaling (solid line in Fig. 4B in the main paper). This
protects the mother cell, where the concentration of IIE$\cdot$AA-P
actually decreases slightly, from spontaneous $\sigma^F$ release.

\subsection*{Critical parameters that determine the behaviour of
the regulatory network} The impact of the individual kinetic
parameters on the behaviour of the model can be assessed by means of
control coefficients, as previously suggested for the analysis of
signaling networks by Lee and co-workers \cite{Lee03_Heinrich}.
Control coefficients can be used to quantify the relative change in
a model variable in response to (small) relative changes in a
parameter value $k_i$. Here they were determined from

\begin{equation} C_i^{RNApol(\sigma^F)} =\frac{\delta
RNApol(\sigma^F)}{RNApol(\sigma^F)}\frac{k_i}{\delta k_i},
\end{equation}

allowing the impact of parameter perturbations on the concentration
of RNApol($\sigma^F$) complexes to be ascertained. To understand the
relative impact of the different interactions we studied the 30
independent rate constants listed in Tables
\ref{Table_par}$\&$\ref{Table_par_invivo}. Control coefficients were
determined in the equilibrium state before septation. We used 15
$\mu$M $\sigma^F$, $30$ $\mu$M AA, $22.5$ $\mu$M AB, and $15$ $\mu$M
IIE for the concentrations of the sporulation proteins, which
reflect the experimentally observed ratios between them and are
within the experimental concentration range
\cite{Magnin97,Lord99,Lucet99}. As required by the summation theorem
for concentration control \cite{Heinrich74} they sum to zero.
Several reactions exert strong control ($C_i^{RNApol(\sigma^F)} \geq
1$), most notably AA phosphorylation and dephosphorylation, but the
reactions responsible for ADP-ATP exchange and the binding of the
transcription factors also have a strong effect (Tables
\ref{Table_par}$\&$\ref{Table_par_invivo}).

Affinities of the transcription factor are particularly interesting
since the relative affinities of $\sigma^F$ and $\sigma^A$ appeared
counterintuitive when first determined \cite{Lord99}. The
mathematical model now reveals that an equal, or higher, affinity of
$\sigma^F$ relative to $\sigma^A$ for the RNA polymerase would
result in a strong increase in RNApol($\sigma^F$) complex formation
even before septation, which could lead to premature activation
(Fig. 3D in the main paper).

We conclude that several safeguards are employed to avoid premature
activation of sporulation notwithstanding the high sensitivity of
the regulatory system to small perturbations in protein
concentration. These safeguards include a tight control of protein
concentrations, an adjustment of the phosphatase rate, the different
affinities of the RNA polymerase for the two transcription factors
and the allosteric behaviour of AB.

\section*{Discussion}
Complex biological networks are difficult to understand from verbal
descriptions, and mathematical modelling is a promising tool to
integrate the available information. However, realistic models
cannot be built without quantitative experimental data. The system
studied here offered several unusual advantages in that the entire
regulatory network could be reconstituted \textit{in vitro} and
specific fluorescence probes were available to study all relevant
interactions. This enabled realistic values for all kinetic
parameters that affect the model to be estimated from experiments,
and allowed the derivation of a very detailed model of the process
that reproduced the experimental data quantitatively. In applying
the model to the physiological situation the phosphatase rate of
IIE, which exhibits a lower rate \textit{in vivo} than \textit{in
vitro} \cite{Feucht02}, was the only parameter that had to be
changed.

We began with a simple, basic circuit from published schemes but
this rapidly grew into a large network after mathematical modeling
predicted that AB is an allosteric protein. Here the combination of
theory and experiment became particularly fruitful because
theoretical predictions led to new experiments and analysis. Some
rate constants, such as the phosphorylation rate and the $\sigma^F
\cdot$AB$\cdot$ADP binding rates, which had been estimated from
previous experiments using simple models, also needed to be
re-evaluated as the regulatory scheme became more complex. Once the
rate and binding constants had been determined, the time-dependent
variation in the system variables (e.g. concentrations of the
pathway components and their complexes) could be analysed by
translating the interactions into a set of ordinary differential
equations, which could be solved numerically. Comparison of the
model predictions to data derived \textit{in vitro} confirmed that
the model was realistic and a perturbation study revealed that the
model is robust to variations in the kinetic parameter values which
justified a direct application of the model to the physiological
situation.

Extrapolation of the model to the physiological situation enabled us
to evaluate many effects that had been associated with the
regulation of $\sigma^F$ release in the prespore and to integrate
them into a mechanistic picture of the regulatory process. We
conclude from analysis of the model that IIE accumulation on the
asymmetrically placed septum is sufficient to trigger
prespore-specific $\sigma^F$ release because IIE association with
its substrate AA-P maximises AA production in the prespore while AB
allostery generates the required sensitivity. Thus, the volume
difference between the two compartments (mother cell and prespore),
AB allostery together with the architecture of the regulatory
network as well as the balanced concentrations of all sporulation
proteins are the key factors that ensure compartmentalized gene
expression during sporulation in \textit{B. subtilis}. Alternative
mechanisms such as a block in AB expression or the removal of an
(elusive) IIE inhibitor, possibly through a transient genetic
imbalance, are unlikely to be of major physiological relevance.

The regulatory system that leads to compartmentalized gene
expression in \textit{B. subtilis} is optimized in many ways. First
of all the effect of the septation-induced difference in the ratio
of surface area to volume is maximized between the two compartments
by concentrating the phosphatase IIE on the asymmetric septum, that
is on the common interface. Given the small size of the bacterium, a
homogeneous distribution of IIE on the surfaces of both compartments
would have led to only a small change in the relative phosphatase
activities. Association of IIE and AA-P ensures  a simultaneous
increase in concentration of both enzyme and substrate in the
prespore, which in turn ensures the strong increase in AA production
that leads to $\sigma^F$ release. Secondly, the protein
concentrations are tuned such that minute changes in IIE and AA can
tip the balance while robustness to fluctuations in gene expression
is ensured by coupled expression combined with protein degradation.
Thirdly, the previously unnoticed allosteric behaviour of the AB
dimer enables the high sensitivity to changes in IIE and AA and at
the same time prevents premature AB$\cdot$ADP$\cdot$AA complex
formation, which would lead to $\sigma^F$ release. Finally, the
formation of AB$\cdot$ADP$\cdot$AA complexes after septation serves
as sophisticated mechanism for preventing $\sigma^F$ from rebinding
to the released AB, thus providing rapid and prolonged $\sigma^F$
release while minimizing cycling and associated wastage of ATP.
AB$\cdot$ADP$\cdot$AA complexes thus serve to remove AB (AB sink
\cite{Lee00,Lee01}) rather than AA (AA sink \cite{Carniol04}).

The model's prediction that pre-septational AB$\cdot$AA complex
formation is detrimental and that AB$\cdot$ADP$\cdot$AA complexe
serve as AA sink stands in direct opposition to the suggestion that
AB$\cdot$ADP$\cdot$AA complexes may act as an AA sink to prevent
premature $\sigma^F$ release \cite{Carniol04}. This suggestion had
been proposed to explain the counterintuitive experimental
observation that unphosphorylated AA can be observed long before
septation but without inducing $\sigma^F$-dependent gene
transcription \cite{King99}. Our model explains the pre-septational
accumulation of (inactive) AA without invoking AB$\cdot$ADP$\cdot$AA
complex formation (Fig. \ref{Fig_phys_prot_express}B,
\ref{Fig_allo}B). The large fraction of AA that is unphosphorylated
at the very onset of sporulation (Fig.\ref{Fig_phys_prot_express}B)
is the consequence of the low protein concentrations which prevent
the formation of protein complexes; lack of AB$\cdot$ATP$\cdot$AA
complex prevents AA phosphorylation. Similarly, most of the
$\sigma^F$ is free at that time, but its concentration is too low to
allow the formation of sufficient $\sigma^F\cdot$RNA polymerase
holoenzyme to effect gene transcription. Once protein concentrations
increase, the model predicts that much AA will be bound in complexes
with AB$\cdot$ATP. Experiments find that half of AA is
unphosphorylated aready before septation, and the simulations
predict that this applies indeed to the soluble AA pool (not IIE
bound).

The critical importance of the simultaneous accumulation of IIE and
its substrate AA-P as well as the need to render septation dependent
on the protein (and in particular IIE) concentration would probably
have remained unnoticed without mathematical modeling. Much the
contrary it had already been reasoned that the release of the
product AA from IIE was inhibited before septation \cite{King99}.

The mathematical model also resolves the paradox of why degradation
of unbound AB is necessary for sporulation while the half-life
($\sim$ 28 minutes) strongly exceeds the time scale on which
$\sigma^F$ needs to be released ($\sim$ 10 minutes)
\cite{Pan01,Hilbert04}. The model reveals that the observed
half-life reduces the AB concentration, whose expression is
presumably coupled with AA \cite{Fort84}, to a level that small
changes in the AA concentration can trigger $\sigma^F$ release.
Coupled expression and degradation of unbound AB (rather than all
AB) confer robustness to fluctuations in gene expression. AB
allostery ensures robustness to parallel changes in protein
concentration (without this feature the initial increase in protein
concentration might cause $\sigma^F$ release in the entire
sporangium). A IIE dependency of septation is likely to ensure that
sufficient protein has been expressed such that septation can
trigger the formation of micromolar $\sigma^F\cdot$RNApol holoenzyme
concentrations. The relative affinities of the transcription factors
$\sigma^F$ and $\sigma^A$ for the core RNA polymerase are tuned to
prevent premature activation whilst enabling the system to be
triggered by small perturbations. Moreover RNApol($\sigma^A$)
holoenzyme remains present when RNApol($\sigma^F$) emerges, in such
a way that general transcription programs can still be carried out.
The robustness of the regulatory circuit to alterations in
concentration and values of kinetic parameters is such that the
process can be deregulated only by mutation.\\

We conclude that the regulatory circuit described here appears to be
optimized to the needs of the organism. It is tuned to ensure both
high sensitivity to stimuli and robustness to random fluctuations.
We show that the dimeric, allosteric kinase SpoIIAB plays a crucial
role by enabling sensitivity, robustness, and minimal ATP
consumption. The high sensitivity to changes in the IIE$\cdot$AA-P
concentration enables the two different developmental programs to be
triggered by small concentration changes (such as can be induced by
asymmetric septation) and with minimal ATP consumption.

While asymmetric cell division, and the resulting difference between
the two compartments, is an elegant mechanism for achieving
compartmentalized gene expression, other symmetry-breaking
mechanisms must exist to induce compartmentalized gene expression in
those sporulating bacteria, such as \textit{Sporosarcina ureae},
that divide symmetrically during sporulation \cite{Zhang97}. A
deeper insight into such alternative mechanisms will be of great
interest.

\clearpage
\newpage

\section*{Appendix: Comments on how to estimate of septation-dependent IIE activity change}
Given that IIE accumulates on the septum (together with any AA-P
that is bound to it) the number of IIE$\cdot$AA-P complexes acting
per unit volume increases in the prespore and decreases in the
mother cell. A lower limit for the septation-dependent increase in
IIE activity in the prespore can be determined by assuming a) that
all the IIE locates, together with its substrate, on to the septum
in such a way that its activity is displayed equally on both faces
and b) that the cell is cuboid.. Insertion of a septum at an xth of
the length of the cuboid will give rise to two compartments, with
the smaller having an xth of the total volume. If half of all
IIE$\cdot$AA-P is present in the smaller volume then there is an
effective $\frac{x}{2}$-fold increase in IIE$\cdot$AA-P per unit
volume of the smaller compartment. Septa form at one-fifth or less
of the length of the cell \cite{Illing91} and the minimal increase
in IIE activity is therefore 2.5.

In fact, each of the assumptions mentioned above underestimates the
probable increase in IIE activity: the total quantity of IIE is
slightly greater on the prespore face than on the mother-cell face
of the septum \cite{Wu98_IIE,King99}; and the prespore is not a
cuboid but has rounded ends \cite{Ryter65}. Thus, although a
2.5-fold increase in IIE activity is sufficient to ensure the
formation of enough $\sigma^F \cdot$RNApol to permit transcription
in the prespore (Fig. \ref{Fig_phys_EAp}C), the 4-fold increase that
we have assumed throughout this paper is a more realistic
representation of the physiological situation.

\newpage

%\bibliography{C:/Science/Texfiles/refs}
%\bibliographystyle{C:/Programme/texmf/bibtex/bst/jtb/jtb}
%\bibliographystyle{C:/Programme/texmf/bibtex/bst/plos/plos_DI_1}
%\bibliographystyle{C:/Programme/texmf/bibtex/bst/plos/plos_numb}

\paragraph{Acknowledgements}
We thank J-C Shu for providing unpublished SPR data. This work was
supported by the BBSRC UK. D.I. is a Junior Research Fellow at St
John's College, University of Oxford and is supported by an EPSRC
scholarship. I.D.C. acknowledges financial support from the Wellcome
Trust and the NIH funded Cell Migration Consortium.

\clearpage
\newpage

\section*{Figures}
\paragraph{Note}
A figure with a scheme of the mathematical model will be included
with the published version of this paper.

\clearpage

\clearpage
\newpage

\clearpage
\newpage

\section*{Tables}
\begin{table}[!h] \caption{\label{Table_par} \textbf{Control coefficients for the kinetic rate constants employed in the \textit{in
vitro} and  \textit{in vivo} simulations.} The determination of
control coefficients $C_j^{RNApol(\sigma^F)}$ is described in the
main text. References to a particular figure show that a parameter
value was derived from results presented in that figure rather than
from previously published results.}
\begin{footnotesize}
\begin{tabular}{|l|c|}
\hline
  \textbf{rates} & \textbf{$C_j^{RNApol(\sigma^F)}$}  \\ \hline
  AA-AB on-rate & $2 \times 10^{-5}$ \\ \hline
  AA-AB off-rate & -0.07  \\ \hline
  AA-$\hat{AB}$.ATP off-rate& 0.03\\ \hline
  AA-$\hat{AB}$.ADP off-rate & $-0.02$  \\ \hline
  AA-$\hat{AB}$ off-rate & $5 \times 10^{-7}$ \\ \hline
  $\sigma^F$-AB on-rate & -0.66  \\ \hline
  $\sigma^F$-AB$\cdot$ATP off-rate & 0.29  \\ \hline
  $\sigma^F$-AB$\cdot$ADP off-rate & $10^{-3}$ \\ \hline
  $\sigma^F$-AB$\cdot$ off-rate & -$3\times 10^{-4}$ \\ \hline
  $\sigma^F$-AB.ATP$\cdot$AA off-rate & 0.15 \\\hline
  $\sigma^F$-AB.ADP$\cdot$AA off-rate & $10^{-6}$ \\\hline
  AB conformational change & $-0.84$ \\ \hline
  AB conformational change & $0.08$\\ \hline
  AB conformational change & $-0.08$\\ \hline
  lid opening T state & -1.16  \\ \hline
  lid opening R state & -$9\times 10^{-6}$  \\ \hline
  lid closure T state &  1.14 \\ \hline
  lid closure R state &  $4\times 10^{-7}$  \\ \hline
  lid closure R state ($\sigma^F$ bound) & $3\times 10^{-4}$\\ \hline
  nucleotide-AB on-rate & -$5\times 10^{-4}$ \\ \hline
  nucleotide-AB off-rate & -0.32\\ \hline
  ADP-AB off-rate (T state) & -$3\times 10^{-3}$  \\ \hline
  ATP-ADP exchange for AB$\cdot$AA& -$6\times 10^{-3}$  \\ \hline
  AA phosphorylation &-1.79  \\ \hline
  AA-P$\cdot$IIE on-rate & 0.035\\ \hline
  AA-P$\cdot$IIE off-rate & -0.034  \\ \hline
  AA dephosphorylation & see Table \ref{Table_par_invivo}  \\ \hline
\end{tabular}
\end{footnotesize}
\end{table}

\begin{table}[!h] \caption{\label{Table_par_invivo} \textbf{Kinetic rate constants only employed in the \textit{in vivo} simulation.}
The determination of control coefficients $C_j^{RNApol(\sigma^F)}$
is described in the main text.}
\begin{footnotesize}
\begin{tabular}{|l|c|c|c|}
\hline
  \textbf{rates} & \textbf{$C_j^{RNApol(\sigma^F)}$} &\textbf{simulation value}  & \textbf{Ref} \\ \hline
  AA dephosphorylation & 3.24 & $3.2\times 10^{-3}$ &  Fig. \ref{Fig_phys_EAp}A, \cite{Feucht02} \\ \hline
  $\sigma$-RNApol on-rate & 0.33 & $10^6$  & \cite{Lord99} \\ \hline
  $\sigma^F$-RNApol off-rate& -0.78& $0.55$  & \cite{Lord99} \\ \hline
  $\sigma^A$-RNApol off-rate& 0.45& $0.02$   & \cite{Lord99} \\ \hline
  AB degradation & n/a & $4.1 \times 10^{-4}$  & \cite{Pan01} \\ \hline
\end{tabular}
\end{footnotesize}
\end{table}

\end{document}